\documentclass[intlimits,twoside,a4paper]{article}

\usepackage[cp1251]{inputenc}
\usepackage{color}
\usepackage{ulem}

\usepackage[eqsecnum]{cmpj3}
\issue{2024}{27}{1}{13801}
\doinumber{10.5488/CMP.27.13801}
\title[Active microrheology of fluids with orientational order]%
{Active microrheology of fluids with orientational order%
}
\author[J. S. Lintuvuori, A. W\"urger] {J. S. Lintuvuori\orcid{0000-0003-4108-9550}, A. W\"urger\thanks{\email{juho.lintuvuori@u-bordeaux.fr},~\email{alois.wurger@u-bordeaux.fr}}}
\address{Universit\'{e} de Bordeaux, CNRS, LOMA, UMR 5798, F-33400 Talence, France}
\Keywords{liquid crystals, colloids, microrheology, lattice Boltzmann methods}
\date{Received July 19, 2023, in final form November 9, 2023}
\begin{document}

\maketitle

\begin{abstract}
We study the dynamics of a driven spherical colloidal particle moving in a fluid with a broken rotational symmetry. Using a nematic liquid crystal as a model, we demonstrate that when the applied force is not aligned along or perpendicular to the orientational order, the colloidal velocity does not align with the force, but forms an angle with respect to the pulling direction. This leads to blue an anisotropic hydrodynamic drag tensor which depends on the material parameters. In the case of nematic liquid crystal, we give  an analytical expression and discuss the resulting implications for  active microrheology experiments on fluids with broken rotational symmetry.
%
%
%\keywords liquid crystals, colloids, microrheology, lattice Boltzmann methods % Up to six keywords (\href{https://physh.aps.org/browse}{Physics Subject Headings})
\printkeywords
%
%\pacs Up to six PACS numbers (optional)
\end{abstract}

\section{Introduction}  

Understanding the flow properties of complex fluids is a challenging goal. Various techniques have been developed to probe the hydrodynamic response of the fluids to external stimuli. An example of this  is provided by shear rheology, which allows the measurement of the bulk viscosity $\eta$. Another modern technique is called microrheology, which consists of measuring the hydrodynamic drag experienced by a spherical colloidal particle  moving in the fluid  due to external~\cite{Squires} or thermal~\cite{manon} forces.  In the case of an external force, the resulting force-velocity curves can be analyzed by using Stokes' law~\cite{landau}: a particle of size $R$ moving at velocity $v$ in a  fluid of viscosity $\eta$ experiences an hydrodynamic drag proportional to these quantities, $-\xi v$ with $\xi=6\piup\eta R$, which balances the external force $F$.

The situation is markedly different when the host fluid has a broken rotational symmetry, such as the orientational order observed in liquid crystals~\cite{degennes}, sheared polymer solutions~\cite{polymer1,polymer2,polymer3} and various biological fluids~\cite{bio1,bio2,bio3}.  Then, the viscous response is described by a fourth-order tensor $\eta_{ijkl}$ acting on the stress components $\sigma_{kl}$~\cite{degennes}, and the resulting hydrodynamic drag force on a spherical bead does not necessarily align with the velocity $\mathbf{v}$ of the particle~\cite{StarkReview}. In other words, the drag coefficient $\xi_i$ becomes a second-order tensor. %where the hydrodynamic drag depends from the angle $\theta$ between the external force $\mathbf{F}$ and the direction of the orientational order $\mathbf{n}$.  

A typical example of fluids with orientational order is provided by thermotropic liquid crystals in a nematic state. Here, the molecules are, on average, aligned along a common direction, characterised by a vector field  $\hat{\mathbf{n}}$ with a head-to-tail symmetry. Recent experimental, theoretical and simulations studies have quantified the drag coefficient $\xi_i=6\piup\eta_i R$ of a spherical colloid, parallel and perpendicular to $\hat{\mathbf{n}}$, and observed $\frac{\xi_{\perp}}{\xi_{||}}=\frac{\eta_{\perp}}{\eta_{||}}\sim 2$~\cite{Terentjev1996,Stark2001,StarkReview,Poulin2004,TiffanyLC,Juho1,Mondiot2012}; this relation also defines effective viscosities $\eta_\perp$ and $\eta_\parallel$.

Here, we study a general case of a spherical particle (radius $R$) dragged through a nematic liquid crystal (LC) with an angle $\theta$ between the pulling force $\mathbf{F}$ and the nematic director $\hat{\mathbf{n}}$. We show that when the pulling direction is between the limiting case of parallel ($\theta = 0^\circ$) and perpendicular ($\theta = 90^\circ$) to the nematic direction, the force and velocity do not align, but a sliding motion is observed. This is characterised by an angle $\alpha$ between the vectors $\mathbf{F}$ and $\mathbf{v}$.

Interpreting this by using the general force-velocity relation via Stokes' law leads to an effective drag response (and thus viscosity) which depends both on the $\theta$ and the ratio of the drag coefficients parallel and perpendicular to the nematic director. We give  analytical expressions for both the generalized drag coefficient $\xi_{\mathrm{eff}} =-F/v$  and the sliding angle $\alpha$, and validate them  by lattice Boltzmann simulations.

\section{Methods}
The nematic order is described by a symmetric and traceless tensor order parameter $\mathbf{Q}$, which time evolution follows a hydrodynamic equation~\cite{berisedwards}
\begin{equation}
(\partial_t + u_{\nu}\partial_{\nu})Q_{\alpha \beta} - S_{\alpha\beta}= \Gamma H_{\alpha \beta},
\end{equation}
where the first part describes the advection and $S_{\alpha\beta}$ describes the possible rotation/stretching of $\mathbf{Q}$ by the flow~\cite{berisedwards}. $\Gamma$ is the rotational diffusion constant. The molecular field is
\begin{equation}
H_{\alpha \beta}= -{\delta 
  {\cal F} / \delta Q_{\alpha \beta}} + (\delta_{\alpha \beta}/3) {\mbox {\rm Tr}}({\delta {\cal F} / \delta Q_{\alpha \beta}}),
\end{equation}
where ${\cal F}$ is  a Landau-de Gennes free-energy whose density can be expressed in terms of a symmetric and traceless order parameter tensor $\mathbf{Q}$ as ${\cal F} = F(Q_{\alpha\beta}) + \tfrac{K}{2}(\partial_{\beta}Q_{\alpha \beta})^2$, with \begin{equation}\label{eq:fed_bulk}
F(Q_{\alpha\beta}) = A_0\left(1-\frac{\gamma}{3}\right)\frac{Q_{\alpha \beta}^2}{2}-\frac{\gamma}{3}Q_{\alpha \beta}Q_{\beta \gamma}Q_{\gamma \alpha} + \frac{\gamma}{4}(Q_{\alpha \beta}^2)^2,
\end{equation}
where Greek indices denote Cartesian coordinates and summation over repeated indices is implied. $A_0$ is a free energy scale, $\gamma$ is a temperature-like control parameter giving a order/disorder transition at $\gamma\sim 2.7$, and $K$ is an elastic constant. 
The anchoring at the particle surface is modelled by $f_s=W(Q_{\alpha\beta}-Q^{0}_{\alpha\beta})^2$, where $W$ is the anchoring strength and $Q^{0}_{\alpha\beta}$ is the preferred alignment of the nematic director at the particle surface.

\begin{figure}[!t]
	\center\includegraphics[width=0.4\columnwidth]{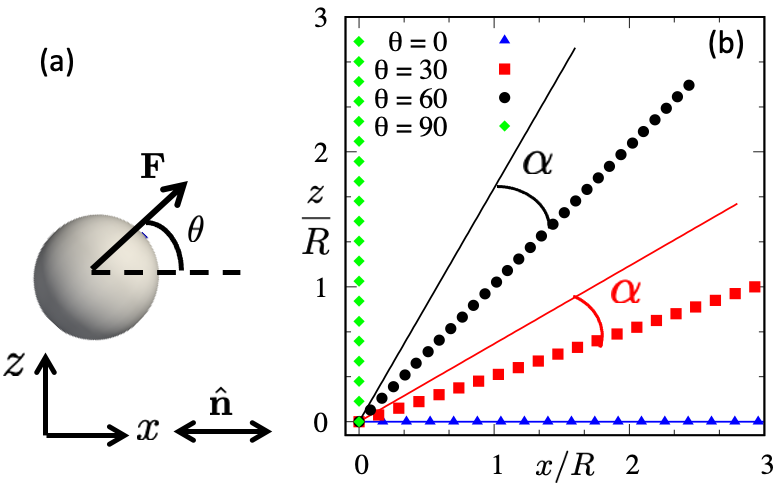}
	\caption{(Colour online) (a) Schematic of the system. (b) The observed steady state trajectories of the particles in the $x-z$ space (symbols) in the absence of surface anchoring ($WR/K=0$). The solid lines show the direction of the applied force. When $0^\circ<\theta<90^\circ$, the force and velocity are not aligned, but the particles move on a straight trajectory with an angle $\alpha$ with respect to the applied force.}
	\label{Schematic}
\end{figure}
%and $\Gamma$ is a collective rotational diffusivity.
The fluid velocity obeys the continuity $\partial_\alpha u_\alpha = 0$, and the Navier-Stokes equations. These are coupled to the LC via a stress tensor. We employ a 3D lattice Boltzmann algorithm to solve the equations of motion (for further details see e.g.~\cite{Juho1,LBLC}).

The colloids are modelled as spherical particles, with a no-slip boundary condition. The no-slip boundary condition at the fluid/solid interface is realized by a standard method of bounce-back on links (BBL)~\cite{ladd1, ladd2} and it can be modified to take into account the moving particle surface~\cite{ladd3}.

\begin{figure}[!t]
	\center \includegraphics[width=0.5\columnwidth]{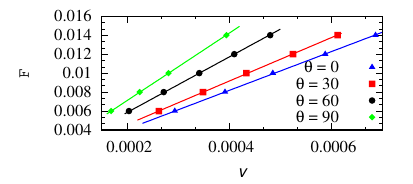}
	\caption{(Colour online) An example of the active microrheological measurement of particles with no surface anchoring ($WR/K = 0$), corresponding to the sample in figure~\ref{Schematic}. The particles were pulled in the nematic LC with a force $F$ at an angle $\theta$ with respect to the nematic director $\hat{\mathbf{n}}$, and the magnitude $v$ of the  velocity $\mathbf{v}$ was measured. The solid lines are linear fits $F=\xi v$.}
	\label{force_v_velocity}
\end{figure}

The dynamics of our system is governed by the Reynolds (Re) and Ericksen (Er) numbers, measuring the ratios between inertial and viscous forces as well as viscous and elastic forces, respectively. Using our simulation parameters~\footnote{Parameters were (in simulation units): 
	$A_0=0.1,~K \simeq 0.01,~\xi=0.7,~\gamma = 3.0$, $q=1/2$ , $\Gamma =0.3$, $W=0$ or $0.01$ and a particle radius $R=4.0$. These give $\tau=0$,  and $\gamma_1=\frac{2q^2}{\Gamma}=\frac{5}{3}$. A  
	cubic simulation box { $16R\times 16R\times 16R$ with periodic boundary conditions}, was used.}, we get the upper limits of $\mathrm{Re}=\frac{vR}{\eta}\approx 0.03$ and $\mathrm{Er} = \frac{\gamma_1 vR}{K}\approx 1.5$, where $\gamma_1$ is the rotational viscosity of the nematic LC.
%~\cite{simulation-units}.

\section{Results}
\begin{figure}
	\center\includegraphics[width=0.5\columnwidth]{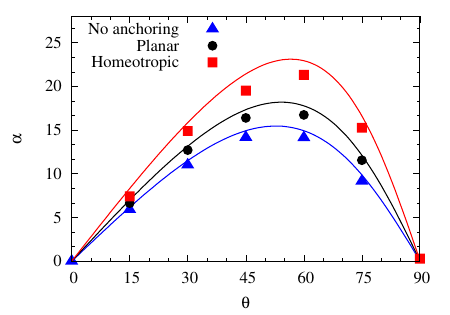}
	\caption{(Colour online) The difference between the sliding angle $\alpha$ and the pulling angle $\theta$ as a function of $\theta$. The symbols correspond to LB simulations and the solid lines are theoretical predictions (equation~\ref{at2}) calculated using the ratio $\frac{\xi_{\perp}}{\xi_{||}}$ from figure~\ref{viscosity}.} 
	\label{alpha}
\end{figure}

 We study the hydrodynamic drag of a spherical particle pulled through a nematic liquid crystal with a force $\mathbf{F}$ on an angle $\theta$ with respect to the nematic director $\hat{\mathbf{n}}$ (figure~\ref{Schematic}a). In addition to the situation where particle surface does not influence the orientation of the nearby liquid crystal molecules ($WR/K=0$), we also consider  a strong  homeotropic or degenerate planar anchoring conditions ($WR/K=4$), where the director adopts either normal or planar orientation with respect to the particle surface, giving a rise to Saturn ring or boojum defects, respectively.

\begin{figure}
	\center\includegraphics[width=0.5\columnwidth]{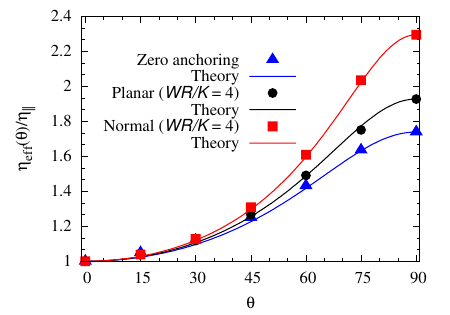}
	\caption{(Colour online) The observed normalized effective viscosity $\frac{\eta_{\mathrm{eff}}}{\eta_{||}}$ as a function of $\theta$ from the simulations (symbols). The solid lines are the theoretical predictions (equation~\ref{drag}) in the text plotted using $\xi_{||}$ and $\xi_\perp$ from the LB simulations.} 
	\label{viscosity}
\end{figure}

When the force is aligned along ($\theta = 0^\circ$) or perpendicular ($\theta=90^\circ$) to $\hat{\mathbf{n}}$, the particle moves along the force with a speed $v$. This allows for a microrheological measurement of the drag coefficients along and perpendicular to the nematic order, where $\frac{\xi_{\perp}}{\xi_{||}}=\frac{\eta_{\perp}}{\eta_{||}}\sim 2$ is observed (figure~\ref{force_v_velocity} and~\ref{viscosity}) in agreement with the previous studies~\cite{Terentjev1996, Stark2001, Poulin2004, TiffanyLC,Juho1,Mondiot2012}.

In the general case, when the external force makes an angle $\theta$ with the nematic director $\mathbf{F}\cdot\hat{\mathbf{n}} = F\cos\theta$, the velocity components parallel and perpendicular to the director satisfy the relations
\begin{equation}
\xi_{||}v_{||} = F\cos\theta ,\quad \xi_{\perp}v_{\perp} = F\sin\theta.
\end{equation}
Unless the drag coefficients are equal $\xi_{||}=\xi_{\perp}$, the velocity is not parallel to the force, but forms an angle $\alpha$ with the pulling direction (figure~\ref{Schematic}b), leading to a sliding motion towards the lower viscosity direction ($x$-axis in our case). 

The force-velocity relation is linear (figure~\ref{force_v_velocity}), which is a minimum requirement for an active microrheological measurement. 
From a simple geometrical considerations, one finds
\begin{equation}
	\tan\left(\theta-\alpha\right) = \frac{v_{\perp}}{v_{||}}=\frac{\xi_{||}}{\xi_{\perp}}\tan\theta
	\label{at1}
\end{equation}
or
\begin{equation}
\alpha = \theta -\arctan\left(\frac{\xi_{||}}{\xi_{\perp}}\tan\theta\right),
\label{at2}
\end{equation}
which agree very well with the simulations  (figure~\ref{alpha}).

The linear velocity $v=\sqrt{v^2_{||} + v^2_{\perp}}$ defines the effective drag coefficient through     
  \begin{equation}
  \frac{1}{\xi_{\mathrm{eff}}(\theta)} = \frac{v}{F} 
           = \sqrt{ \frac{ \cos^2\theta}{\xi^{2}_{\perp}} + \frac{\sin^2\theta}{ \xi^{2}_{||}}},
\label{drag}
\end{equation}
and thus, an angle dependent effective viscosity $\eta_{\mathrm{eff}}=\xi_{\mathrm{eff}}/6\piup R$.

\section{Conclusions}

Using the lattice Boltzmann simulations, we carried out microrheological experiments of colloidal particles in a nematic liquid crystal. Both the angle $\theta$ and the surface properties of the particles were varied. In all the cases, we observed a linear relation between the magnitude $F$ of the imposed force and the observed speed $v$. The drag coefficients are trivially available from   linear fits $F=\xi v$ (see  e.g. figure~\ref{force_v_velocity}). The simulations confirmed that the sliding motion towards the lower viscosity direction, leads to a non-trivial flow response (figure~\ref{viscosity}). As can be expected from the sliding angle (figure~\ref{alpha}), the effect is most pronounced for colloids with a strong homeotropic anchoring at their surface followed by  particles with a planar surface anchoring. This can be understood in terms of the strength of coupling between the particle surface and the local nematic order. The effective drag coefficients follow, to a very high degree, the theoretical predictions of equation (\ref{drag}), for all the three surface anchorings considered (figure~\ref{viscosity}).

\ukrainianpart

\title{Активна мікрореологія плинів з орієнтаційним порядком}
\author[Й. С. Лінтувуорі, А. Вюргер] {Й. С. Лінтувуорі, А. Вюргер}
\address{Університет Бордо, CNRS, UMR 5798, F-33400 Таленс, Франція}

\makeukrtitle

\begin{abstract}
	Досліджується динаміка керованої сферичної колоїдної частинки, що рухається в рідині з порушеною обертальною симетрією. Використовуючи в якості моделі нематичний рідкий кристал, показано, що коли прикладена сила не діє вздовж або перпендикулярно орієнтаційному порядку, швидкість частинки не є колінеарною з напрямком сили, а утворює кут відносно напрямку розтягування. Це призводить до ``синього'' анізотропного гідродинамічного тензора опору, який залежить від параметрів матеріалу. У випадку нематичного рідкого кристала ми наводимо його аналітичний вираз і обговорюємо наслідки для активних мікрореологічних експериментів на рідинах із порушеною обертальною симетрією.
	\keywords рідкі кристали, колоїди, мікрореологія, ґраткові методи Больцмана 
\end{abstract}

  \end{document}